\documentclass[sensors,accept,communication,moreauthors,pdftex]{./mdpi}

\firstpage{1} 
\pubvolume{1}
\issuenum{1}
\articlenumber{0}
\pubyear{2023}
\copyrightyear{2023}
\datereceived{ } 
\daterevised{ } 
\dateaccepted{ } 
\datepublished{ } 
\hreflink{https://doi.org/} 

\Title{Single-Element Dual-Interferometer for Precision Inertial Sensing: Sub-picometer Structural Stability and Performance as a Reference for Laser Frequency Stabilization}

\TitleCitation{Title}

\Author{Victor Huarcaya $^{1}$\orcidD{}, Miguel~Dovale~\'Alvarez $^{1,}$*\orcidC{}, Kohei Yamamoto $^{1}$\orcidB{}, Yichao Yang $^{1,2}$\orcidA{}, Stefano Gozzo $^{1}$\orcidE{}, Pablo Martinez Cano $^{1}$\orcidH{}, Moritz~Mehmet~$^{1,3}$\orcidJ{}, Juan Jose Esteban Delgado $^{1}$\orcidI{}, Jianjun Jia $^{4,5}$, and~Gerhard Heinzel $^{1}$\orcidK{} } 
\AuthorNames{Victor Huarcaya, Yichao Yang, Kohei Yamamoto, Juan Jose Esteban Delgado, Moritz Mehmet, Jianjun Jia, Gerhard Heinzel, Miguel Dovale \'Alvarez}

\address{%
$^{1}$ \quad Max-Planck-Institut f\"ur Gravitationsphysik (Albert-Einstein-Institut)  and Institut f\"ur Gravitationsphysik, Leibniz Universit\"at Hannover, Callinstrasse 38, D-30167 Hannover, Germany; victor.huarcaya@aei.mpg.de (V.H.); miguel.dovale@aei.mpg.de (M.D.A); kohei.yamamoto@aei.mpg.de (K.Y.); yichao.yang@aei.mpg.de~(Y.Y.); stefano.gozzo@aei.mpg.de (S.G.); pablo.martinez.cano@aei.mpg.de (P.M.C); moritz.mehmet@aei.mpg.de (M.M.); juan.jose.esteban@aei.mpg.de (J.J.E.D.); gerhard.heinzel@aei.mpg.de~(G.H.)\\
$^{2}$ \quad Laser Link (Shanghai) Aerospace Technology Co., Ltd, Shanghai, China; badunjiangjund@163.com \\
$^{3}$ \quad Texas A\&M University, Department of Physics \& Astronomy, College Station, Texas 77843, USA \\
$^{4}$ \quad Key Laboratory of Space Active Opto-electronics Technology, Shanghai Institute of Technical Physics, Chinese Academy of Sciences, Shanghai 200083, China; jjjun10@mail.sitp.ac.cn \\
$^{5}$ \quad University of Chinese Academy of Sciences, Beijing 100049, China

}
\corres{Correspondence: miguel.dovale@aei.mpg.de}

\abstract{To reach sub-picometer sensitivity in the millihertz range, displacement sensors based on laser interferometry require suppression of laser-frequency noise by several orders of magnitude. Many optical frequency stabilization methods exist with varying levels of complexity, size, and performance. In this paper, we describe the performance of a compact Mach-Zehnder interferometer based on a monolithic optic. The setup consists of a commercial fiber injector, a custom-designed pentaprism used to split and recombine the laser beam, and two photoreceivers placed at the complementary output ports of the interferometer. The structural stability of the prism is transferred to the laser frequency via amplification, integration, and feedback of the balanced-detection signal, achieving a fractional frequency instability better than 6 parts in $10^{13}$, corresponding to an interferometer pathlength stability better than $10^{-12}$\,m$/\sqrt{\mathrm{Hz}}$.}


\begin{document}


\section{Introduction}

In the realm of high-precision inertial sensing, where tiny displacements can hold profound significance, the pursuit of ever-higher measurement sensitivity has driven innovation in optical metrology. Among these, laser interferometry has become a standard tool, particularly in the field of experimental gravitational physics, where it is used, e.g., to reveal the gravity field of Earth~\cite{GFO2019, rs14133092}, understand climate change~\cite{Tapley2019}, or uncover new astrophysical systems through the observation of gravitational waves~\cite{Abbott2016, Abbott2017, Miller2019}.

The GRACE Follow-On (Gravity Recovery and Climate Experiment Follow-On) mission, launched to orbit on May 22, 2018, measured changes in Earth's gravitational field with unparalleled precision by deploying a pair of identical satellites in low Earth orbit~\cite{Kornfeld2019}. These satellites used microwave and laser ranging instrumentation to detect variations in gravitational forces experienced as they orbited the planet~\cite{Abich2019}. Such variations are indicative of changes in mass distribution, offering insights into water storage, ice melt, and land movement. GRACE Follow-On has significantly contributed to understanding Earth's water cycle, ice sheet dynamics, groundwater depletion, and other essential aspects of the Earth's climate~\cite{Tapley2019, Velicogna2020, Landerer2020}.

The accelerometers onboard each spacecraft are used to measure non-gravitational accelerations on the spacecraft in order to separate their effects from the effects of gravitational accelerations, which are the observables of interest. Non-gravitational forces include residual imbalanced thruster firings, and non-gravitational environmental effects, such as atmospheric drag and solar and Earth radiation pressures.

Each accelerometer houses a test mass (TM) which is kept in a state of nearly free fall. The accelerometer monitors the position and motion of the TM using capacitive sensing by placing electrodes on both the TM and its housing. When the TM is perturbed due to inertial forces acting on it, the gap between the electrodes changes, altering the capacitance of the system. The accelerometer measures this change in capacitance, which can be related to the displacement of the TM with $\sim 100\,\rm pm /\!\sqrt{Hz}$ sensitivity.

The accelerometers are at present a dominant noise source in GRACE-like missions, with stray acceleration at the $\sim 10^{-10}\,\rm ms^{-2}/\!\sqrt{Hz}$ level~\cite{Abich2019, Flechtner2015, Wegener2020}. They could be improved by using technology from the LISA Pathfinder mission~\cite{Armano2016}, which demonstrated $\sim\,10^{-15}\,\rm ms^{-2}$ residual TM acceleration in interplanetary orbit~\cite{Armano2021, Armano2022}, employing laser interferometric readout at the level of $1\,\rm pm/\!\sqrt{Hz}$ displacement noise instead of electrostatic readout at $\sim\,100\,\rm pm/\!\sqrt{Hz}$. A reduction in the TM acceleration noise can lead to an important improvement in the scientific return of future geodesy missions focusing on mass change, especially in a scenario with multiple pairs of geodesy satellites~\cite{Weber2022}.

Laser interferometric inertial sensors aiming to measure TM displacements with sub-picometer precision over time scales of hundreds and even thousands of seconds rely on some form of reduction of laser-frequency noise,  which couples through optical pathlength mismatches between the interfering arms. The most common methods of laser-frequency stabilization are locking to an ultra-stable optical cavity~\cite{Drever1983, Webster2007, Webster2008, Webster2011}, or to an atomic or molecular transition~\cite{Arie1992, Leonhardt2006}. These methods are costly, bulky, and rely on complex lock-acquisition schemes. An alternative technique is locking to a Mach-Zehnder interferometer (MZI) with several centimeters of optical pathlength difference~\cite{Gerberding2017}. By introducing an intentional arm length mismatch between the arms, the interferometer's output signal acquires a sinusoidal dependence on the laser frequency, creating an opportunity for laser locking. Furthermore, the two output ports of the interferometer can be subtracted to derive a signal that conserves the sinusoidal dependence, but is largely insensitive to laser power fluctuations.

On a recent paper~\cite{bib:VH}, laser-frequency stabilization via locking to an unequal-arm MZI was shown to provide a stability similar to that of high-performance reference lasers based on molecular iodine hyperfine transitions. By combining a quasi-monolithic interferometer and quasi-monolithic fiber injector with a very stable thermal environment, a fractional frequency instability on the level of a few parts in $10^{13}$ was achieved for measuring times up to 1000 seconds.

Despite the high performance for a system of that size, the MZI in~\cite{bib:VH} has one major drawback in that its manufacturing and assembly is comparatively complex. The interferometer consists of a custom quasi-monolithic fiber injector, five coated fused-silica components (two mirrors and three beam-splitters), and an ultra-stable glass ceramic baseplate. All of the optical components have to be bonded to the baseplate using UV adhesive in a time-consuming and delicate process. The component alignment is done using a coordinate measurement machine and a combination of template-assisted positioning for uncritical components and an adjustable pointing finger assembly for the critical recombination beam-splitter~\cite{bib:katha}.

The custom fiber injector is itself complex, consisting of five components, four of which are custom-designed parts, that also have to be bonded together using UV adhesive. The fiber injector has to be pre-assembled in another time-consuming and delicate process, before it can be included in the full interferometer assembly.

In this paper, we present an unequal-arm Mach-Zehnder interferometer made of a single optical component that can reach sub-picometer displacement sensitivity at 2\,mHz using a commercial fiber injector, on a package that is several times smaller than the interferometer presented in~\cite{bib:VH}. The construction of the prism can be outsourced to a company specializing in optics manufacturing, and its assembly in the laboratory is straightforward. This interferometer design was first presented in~\cite{bib:SEDI} and its performance is described here.

\section{Experimental setup}

\begin{figure}[t!]
\begin{adjustwidth}{-\extralength}{0cm}
\centering
\includegraphics[width=17cm]{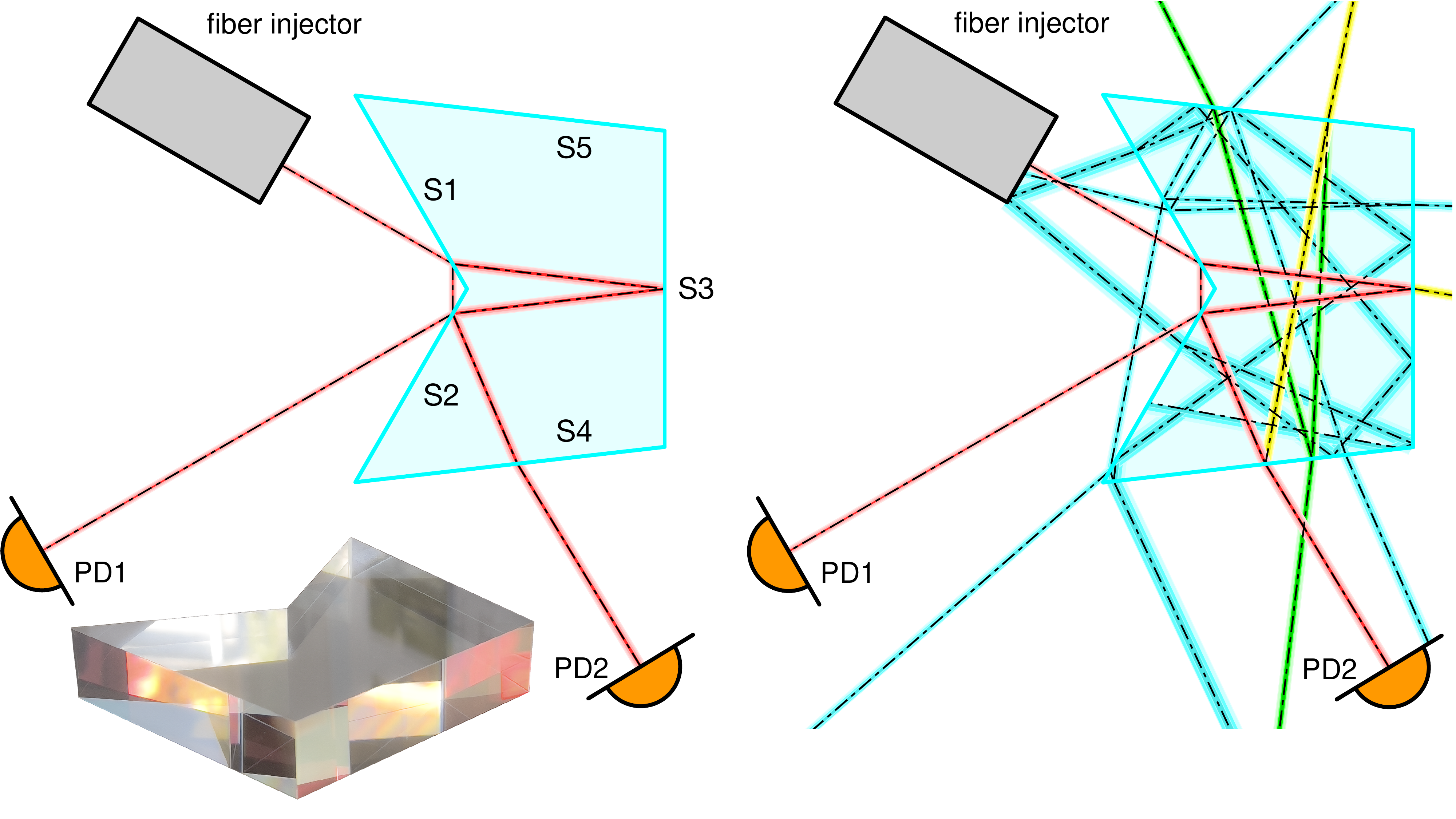}
\end{adjustwidth}
\caption{Single-element dual-interferometer (SEDI) prism showing only the reference interferometer. The prism consists of five optical surfaces with different coatings. Surfaces S1 and S2 split and recombine the beams. Surface S3 acts as a highly reflective mirror for the long arm beam, but is anti-reflective coated on the sides. Surfaces S4 and S5 are anti-reflective coated to minimize the impact of stray light in the setup. The main interferometric beams are shown in red (a). The prism geometry has been optimized via optical simulations to minimize the impact of ghost beams in the photodiodes. Ghost beams with $1 \!>\! P_{\text{ghost}}/P_{\text{main}} \!>\! 10^{-3}$ (yellow), $10^{-3} \!>\!P_{\text{ghost}}/P_{\text{main}} \!>\! 10^{-7}$ (green) and $10^{-7} \!>\! P_{\text{ghost}}/P_{\text{main}} \!>\! 10^{-12}$ (cyan) relative power level are depicted in (b). Two detectors (PD1 and PD2) are placed at the complementary output ports of the interferometer to derive the balanced-detection signal used for laser frequency stabilization. }
\label{figure:optocad}
\end{figure}   

The custom-designed pentaprism is shown in Figure~\ref{figure:optocad}. A monolithic piece of fused silica glass was formed by optical contacting two smaller prisms. A total of six coating runs were applied to the prism. Two 50:50 beam-splitter coatings were applied to surfaces S1 and S2. Surface S3 was first coated anti-reflective (AR) except for a small portion of the surface in the middle, where it is coated high-reflective (HR). Finally, two AR coatings were applied to surfaces S4 and S5.

\begin{figure}[t!]
\begin{adjustwidth}{-\extralength}{0cm}
\centering
\includegraphics[width=17cm]{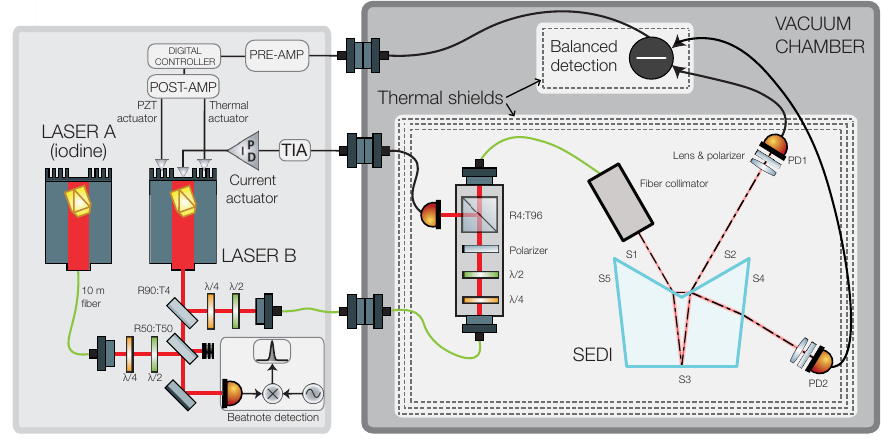}
\end{adjustwidth}
\caption{Experimental setup. The single-element dual-interferometer (SEDI) prism is surrounded by thermal shields and placed inside a vacuum chamber along with auxiliary optics for polarization adjustment and amplitude stabilization. A commercial fiber collimator is used to inject a beam derived from laser B into the SEDI reference interferometer. Two photodiodes are placed at the complementary output ports of the interferometer, with a focusing lens and thin film polarizer placed in front that help mitigate known noise sources. The difference current between the two photodiodes is converted to a voltage via a low-noise low-drift transimpedance amplifier. The amplifier signal is filtered, digitized, and used as input in a digital PI-controller to derive a control signal that is fed back to laser B's fast and slow actuators, thereby transferring the interferometer's pathlength stability to the laser frequency. A beatnote signal in the order of a few GHz is obtained by interfering laser B with a second, more stable laser (laser A). The beatnote signal is mixed-down to below 100\,MHz using an ultra-stable GHz source, and read by a micro-cycle-stable phasemeter to characterize the achieved stability.} 
\label{figure:setup}
\end{figure}  

The prism can form an unequal-arm Mach-Zehnder interferometer for a beam incident on S1 at a certain position and orientation. In the resulting interferometer, the short-arm beam propagates outside of the optic, while the long-arm beam propagates inside with an optical pathlength difference of $l_{\mathrm{ref}} = 144$\,mm between the two. The two beams are recombined at S2 and their interference is captured by the photodiodes PD1 and PD2, placed on the complementary output ports of the recombination beam-splitter. The power at the photodiodes depends on the laser frequency $f$ and is given by
\begin{align}
P_1(f) &= p_1 \left[ 1 + c_1 \cdot \cos \left(\frac{2\pi f \Delta l}{c} + \varphi_0 \right) \right] \nonumber\\
P_2(f) &= p_2 \left[ 1 - c_2 \cdot \cos\left(\frac{2\pi f \Delta l}{c} + \varphi_0  \right) \right]
\end{align}
where $p_{1,2}$ are the optical powers at each photodetector in mid-fringe, $c_{1,2}$ are the interferometric contrasts at each photodetector, $\Delta l$ is the interferometer's optical path length difference, $c$ is the speed of light, and $\varphi_0$ is an arbitrary constant. After a direct current subtraction, and trans-impedance amplification, the resulting signal is given by
\begin{align}
v(f) &= G \left[ P_1(f) - P_2(f) \right] \nonumber\\
&= G \left[ p_1 - p_2  + (c_1 p_1 + c_2 p_2) \cdot \cos \left( \frac{2\pi f \Delta l}{c} + \varphi_0 \right)\right]
\end{align}
where $G\,[\mathrm{V}/\mathrm{W}]$ is the trans-impedance gain. If balanced operation is achieved, with nearly equal power levels on both photodiodes (i.e., $p_1 = p_2 $), the signal becomes
\begin{equation}
v(f) = G p_1 (c_1 + c_2 ) \cdot \cos \left( \frac{2\pi f \Delta l}{c} + \varphi_0 \right)
\label{equation:mzi}
\end{equation} 
Equation~\ref{equation:mzi} has periodic zero crossings that are independent of the laser beam power and are used to lock the laser's frequency. The slope of the error signal at the operating point is proportional to the available optical power, the interferometric contrasts, the trans-impedance gain, and the interferometer's arm length difference.

A focusing lens is placed in front of each photodiode to minimize transverse beam walk, and thin-film polarizers with high extinction ratios are placed after the lens to mitigate the impact of stray light. Balanced operation is obtained by adjusting the polarizer's rotation angles such that both photodiodes receive the same amount of laser power at the mid-fringe operating point.

In addition to the aforementioned interferometer, called ``reference interferometer'' (Ref.\ IFO), the prism was designed to be injected with a second beam, derived from the same laser source, to form an additional interferometer called ``test mass interferometer'' (TM IFO)~\cite{bib:SEDI}. The TM IFO phase would contain the test mass displacement signal which could be recovered by employing a suitable phase readout method~\cite{bib:DFM, bib:DPM}. Due to its characteristics, the prism was dubbed ``single-element dual-interferometer'', or SEDI.

To characterize the performance of the reference interferometer in SEDI, we use the experimental setup depicted in Figure~\ref{figure:setup}. Light from a 1064\,nm non-planar ring oscillator laser (laser B) is split two ways, with one part being fed to a vacuum chamber containing the SEDI prism, and the remaining part being interfered with a reference 1064\,nm laser (laser A) that is locked to a molecular iodine hyperfine transition (R(56)32-0 `a1').

Inside the vacuum chamber, the light is first injected into a small bench where a combination of retarder waveplates and a polarizer produce s-polarized light. A small portion of the light is captured by an auxiliary photodiode and used for stabilization of the laser amplitude, and the rest is coupled back into a fiber and injected into the SEDI prism's Ref.\ IFO via a commercial fiber coupler. All of the aforementioned components are mounted on an aluminum breadboard and surrounded by a high-performance multi-layer thermal shield similar to the one described in~\cite{Dovale2019PhD}.

The photodiodes are operated in reverse bias voltage and connected in a balanced differential trans-impedance amplifier (TIA) performing a direct current subtraction, giving rise to a signal with sinusoidal dependence on the laser frequency and the interferometer's pathlength noise. The TIA is surrounded by a separate thermal shield to avoid temperature cross-couplings between the optics and electronics.

The TIA output signal is low-pass-filtered and enhanced by a pre-amplifier before being digitized by a Moku:Lab instrument~\cite{bib:moku} and used as error signal in a digital PI-controller. The resulting control signal is filtered by a post-amplifier and fed back to the laser via both a slow thermal actuator and a high-speed piezo-electric transducer actuator.

The beatnote between lasers A and B is in the GHz regime, and thus it is down-mixed to below 100\,MHz by an ultra-stable GHz signal generator (SMB100A by Rohde \& Schwarz) before being read out by a Moku:Lab instrument acting as phasemeter.

\section{Results}

\begin{figure}[t!]
\begin{adjustwidth}{-\extralength}{0cm}
\centering
\includegraphics[width=9cm]{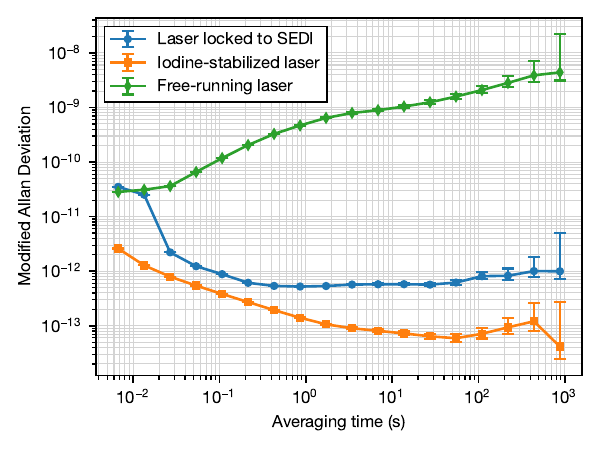}
\includegraphics[width=9cm]{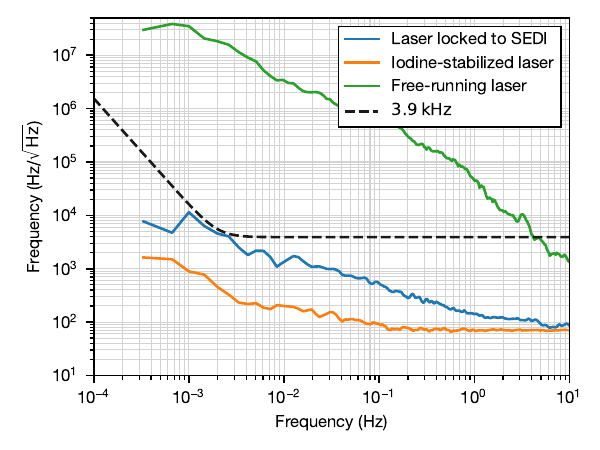}
\end{adjustwidth}
\caption{Modified Allan deviations (a) and frequency spectral densities (b) of laser B when it is free-running (green), laser B when it is locked to the SEDI reference interferometer (blue), and laser A, which is used as reference (orange). The picometer-equivalent frequency noise for a 14.4-cm interferometer is represented by a black dashed curve at 3.9\,kHz$/\!\sqrt{\mathrm{Hz}} \times u(f)$. The error bars in (a) are computed using the ``finite differences'' method~\cite{Greenhall2003}.}
\label{figure:result1}
\end{figure}

The frequency spectral density~\cite{Troebs2006} and modified Allan deviation~\cite{NIST1065} of the beatnote between the laser locked to SEDI and the iodine-stabilized reference laser are shown in Figures~\ref{figure:result1}a and~\ref{figure:result1}b respectively for a typical 1-hour measurement at a rate of 150 samples per second (blue curves). Also shown are the free-running noise of the laser (green), and the noise of the reference laser (orange). Figure~\ref{figure:result1}b also shows the frequency noise spectral density of 3.9\,kHz$/\!\sqrt{\mathrm{Hz}} \times u(f)$ (black dashed line), representing the picometer-equivalent frequency instability of an interferometer with 14.4\,cm arm length difference, as is the case in the SEDI Ref.\ IFO. The noise envelope function
\begin{equation}
u(f) = \sqrt{1+\left(\frac{2\,\mathrm{mHz}}{f}\right)^4}
\end{equation}
is used frequently to scale the sensitivity requirements for inertial sensing of freely-floating test masses in space. It describes a mixture of \emph{white noise} with a flat power spectrum at frequencies larger than 2\,mHz, and \emph{random run} noise with $f^{-4}$ power spectrum at lower frequencies, where it is expected that the acceleration noise of the test mass becomes dominant.

The modified Allan deviation is chosen for its ability to distinguish between white and flicker phase noise at short averaging times (i.e., at short $\tau = m \tau_0$, where $\tau$ is the averaging time, $\tau_0$ is the gate time or sampling time, and $m$ is the averaging factor), or equivalently at high frequencies. This functions is also widely used in the frequency standards community, such that our results may be easily compared with other references. 

Inspection of Figures~\ref{figure:result1}a and~\ref{figure:result1}b, which provide largely the same information, reveals that the noise of the reference laser (laser A) is low enough compared to the laser under test (laser B), that its instability can be neglected in the estimation of the noise of the unit under test. The laser locked to the SEDI prism presents a fractional frequency instability below the $10^{-12}$ level for averaging times between 0.1 and 1000 seconds. This performance is similar to what can be expected from high-performance iodine-stabilized reference systems.

\begin{figure}[t!]
\centering
\includegraphics{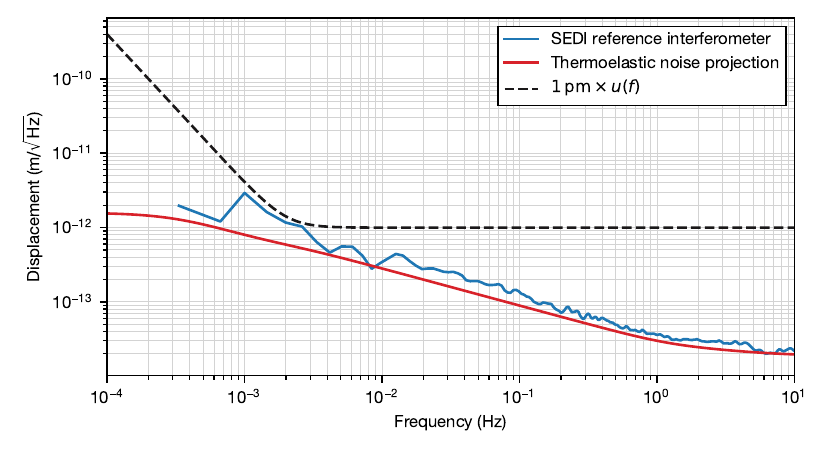}
\caption{Amplitude spectral density of the SEDI reference interferometer pathlength (blue) and projection of the thermoelastic noise (red). A displacement noise of 1\,pm$/\!\sqrt{\mathrm{Hz}} \times u(f)$ is represented by the black dashed curve. }
\label{figure:result2}
\end{figure}

We convert the measured fractional frequency instability into equivalent pathlength noise by invoking
\begin{equation}
\frac{\Delta (l_{\mathrm{ref}})}{l_{\mathrm{ref}}} = \frac{\Delta f}{f_0},
\end{equation}
where $f_0$ is the average laser frequency (roughly 282\,THz). The resulting pathlength noise is shown in Figure~\ref{figure:result2}, together with a projection of thermoelastic noise obtained via numerical modeling, using the model described in~\cite{bib:SEDI} and assuming a uniform temperature distribution of $20\,\upmu\mathrm{K}/\sqrt{\mathrm{Hz}}$ spectral density in the surface of the prism, which is consistent with our measurements. 

The test mass displacement sensitivity of the SEDI test mass interferometer can be estimated by multiplying the measured reference interferometer pathlength noise by the ratio between armlength differences in both interferometers (i.e., approximately 3.5). This results in $10\,\rm pm/\sqrt{Hz}$ sensitivity at 1\,mHz, $1\,\rm pm/\sqrt{Hz}$ sensitivity at 10\,mHz, and $0.1\,\rm pm/\sqrt{Hz}$ sensitivity at 1\,Hz, which is in good agreement with the noise budget analysis presented in~\cite{bib:SEDI}.

A comparison of the projected SEDI test mass displacement noise to previous works on compact interferometric inertial sensors is given in Table~\ref{table:comparison}. For additional references, including earlier works, see the comprehensive overview by Watchi et al~\cite{Watchi2018} on compact interferometers.
 
\begin{table*}
\centering
\begin{adjustwidth}{-1cm}{0cm}
\begin{tabular}{clcccc}
\toprule
 & & Noise at 1\,mHz & Noise at 1\,Hz & Wavelength & Dimensions \\
Year & Device & $(\rm pm/\sqrt{Hz})$ & $(\rm pm/\sqrt{Hz})$ & (nm) & (cm) \\
\midrule
2018 & Pisani~\cite{Pisani2018} & $2 \cdot 10^3$ & $0.6$ & 1064 & $ 11.0 \times 10.0 \times 6.0 $ \\
2019 & Isleif~\cite{Isleif2019}$^{\dagger}$ & $20$ & $0.23$ & 1064 & $ 2.5 \times 2.5 \times 2.5$ \\
2022 & Yan~\cite{Yan2022} & $10^4$ & $4-6$ & 1064 & $ 20.0 \times 20.0 \times 7.0 $ \\
2022 & Zhang~\cite{Zhang2022}  & $10^3$ & $0.6$ & 1550 & $ 2.0 \times 2.0 \times 1.0 $ \\
2022 & Smetana~\cite{Smetana2022} & -\,- & $0.3$ & 1064 & $ 1.3 \times 0.4 $ \\
2022 & Kranzhoff~\cite{Kranzhoff2022} & $10^9$ & $1$ & 1064 & $ 32.0 \times 23.0 \times 31.0 $ \\
2023 & Huarcaya~\cite{bib:VH} & 0.4 & 0.007 & 1064 & $13.5 \times 13.5 \times 7.1$ \\
2023 & SEDI (this work)$^{\dagger\dagger}$ & 10.2 & 0.12 & 1064 & $9.8 \times 7.8 \times 2.0$ \\
\bottomrule
\end{tabular}
\end{adjustwidth}
\caption{Comparison of the sensitivity of compact interferometers in recent works. $^{\dagger}$ In this work, a separate frequency reference interferometer is used to reduce the laser frequency noise. $^{\dagger\dagger}$ The noise is projected from the measured reference interferometer noise floor.}
\label{table:comparison}
\end{table*}

\section{Conclusions}

A Mach-Zehnder interferometer with unequal arm lengths was implemented in a custom pentaprism formed by single piece of fused silica glass. The optic was constructed by optical contacting two smaller prisms and applying six coating runs to the five optical surfaces. The interferometer's balanced-detection signal was used to lock the frequency of a laser down to a fractional instability below $10^{-12}$ for averaging times between 0.1 and 1000 seconds. The equivalent pathlength stability of the interferometer is better than 1\,pm$/\!\sqrt{\mathrm{Hz}} \times u(f)$ from 1\,mHz to 10\,Hz.

This paper presents the first experimental demonstration of the SEDI optic, whose design was introduced in~\cite{bib:SEDI}. The performance characterization of the reference interferometer in SEDI is vital before attempting the next step: the dual-interferometer configuration. From the noise model described in~\cite{bib:SEDI}, and assuming that the Ref.\ IFO and TM IFO phase noises are uncorrelated, we expect the upper bound of the Ref.\ IFO contribution to the test mass displacement noise to be 10\,pm at 1\,mHz and 1\,pm at 10\,mHz. This contribution is dominated by thermoelastic deformation. The actual noise contribution may be lower as we expect there to be some level of coherence in the thermal noise of the two interferometers.

The advantage of the single-element interferometer over the multi-element approach is the ease of manufacture and assembly. While the Mach-Zehnder interferometer presented in~\cite{bib:VH} requires specialized assembly procedures, the SEDI prism does not. We expect that the difference in their performance comes down to thermoelastic noise, which could be lower in~\cite{bib:VH} due to the use of the ultra-low-expansion glass ceramic baseplate.

Unequal-arm Mach-Zehnder interferometers are an attractive solution for laser-frequency stabilization in a compact setup. In contrast to cavity-locking and atomic or molecular references, the technique offers a wide operating range and does not rely on complex lock acquisition procedures. Since the Mach-Zehnder can be integrated as part of the optical bench in future gravity missions already featuring an optical bench assembly, it holds the potential to eliminate the need for a separate laser stabilization subsystem.

\vspace{6pt} 

\authorcontributions{For research articles with several authors, a short paragraph specifying their individual contributions must be provided. The following statements should be used ``Conceptualization, M.D.A., K.Y., and Y.Y.; methodology, V.H., M.D.A., P.M.C., and M.M.; formal analysis, M.D.A. and S.G.; investigation, V.H. and S.G.; resources, J.J.E.D., J.J., and G.H.; writing---original draft preparation, V.H. and M.D.A.; writing---review and editing, V.H. and M.D.A.; visualization, V.H. and M.D.A.; supervision, M.D.A.; project administration, J.J.E.D., J.J., and G.H.; funding acquisition, J.J.E.D., J.J., and G.H.; All authors have read and agreed to the published version of the manuscript.}

\funding{This work was funded by the Deutsche Forschungsgemeinschaft (DFG, German Research Foundation) Project-ID 434617780-SFB 1464. The authors acknowledge support from the Deutsche Forschungsgemeinschaft (DFG) Sonderforschungsbereich 1128 Relativistic Geodesy and Cluster of Excellence "QuantumFrontiers: Light and Matter at the Quantum Frontier: Foundations and Applications in Metrology" (EXC-2123, Project No.\ 390837967) and Max Planck Society (MPS) through the LEGACY cooperation on low-frequency gravitational wave astronomy (M.IF.A.QOP18098). 

The authors also acknowledge support by the German Aerospace Center (DLR) with funds from the Federal Ministry of Economics and Technology (BMWi) according to a decision of the German Federal Parliament (Grant No.\ 50OQ2301, based on Grants No.\ 50OQ0601, No.\ 50OQ1301, No.\ 50OQ1801).}

\conflictsofinterest{The authors declare no conflict of interest.} 
\appendixtitles{no} 
\reftitle{References}

\begin{thebibliography}{999}

\bibitem[Kornfeld et~al.(2019)Kornfeld, Arnold, Gross, Dahya, Klipstein, Gath,
  and Bettadpur]{GFO2019}
Kornfeld, R.P.; Arnold, B.W.; Gross, M.A.; Dahya, N.T.; Klipstein, W.M.; Gath,
  P.F.; Bettadpur, S.
\newblock GRACE-FO: the gravity recovery and climate experiment follow-on
  mission.
\newblock {\em Journal of spacecraft and rockets} {\bf 2019}, {\em
  56},~931--951.
\newblock {\url{https://doi.org/10.2514/1.A34326}}.

\bibitem[Weber et~al.(2022)Weber, Bortoluzzi, Bosetti, Consolini, Dolesi, and
  Vitale]{rs14133092}
Weber, W.J.; Bortoluzzi, D.; Bosetti, P.; Consolini, G.; Dolesi, R.; Vitale, S.
\newblock Application of LISA Gravitational Reference Sensor Hardware to Future
  Intersatellite Geodesy Missions.
\newblock {\em Remote Sensing} {\bf 2022}, {\em 14}.
\newblock {\url{https://doi.org/10.3390/rs14133092}}.

\bibitem[Tapley et~al.(2019)Tapley, Watkins, Flechtner, Reigber, Bettadpur,
  Rodell, Sasgen, Famiglietti, Landerer, Chambers, Reager, Gardner, Save,
  Ivins, Swenson, Boening, Dahle, Wiese, Dobslaw, Tamisiea, and
  Velicogna]{Tapley2019}
Tapley, B.D.; Watkins, M.M.; Flechtner, F.; Reigber, C.; Bettadpur, S.; Rodell,
  M.; Sasgen, I.; Famiglietti, J.S.; Landerer, F.W.; Chambers, D.P.;  et~al.
\newblock Contributions of {GRACE} to understanding climate change.
\newblock {\em Nature Climate Change} {\bf 2019}, {\em 9},~358--369.
\newblock {\url{https://doi.org/10.1038/s41558-019-0456-2}}.

\bibitem[Abbott et~al.(2016)Abbott, Abbott, Abbott, Abernathy, Acernese,
  Ackley, Adams, Adams, Addesso, Adhikari, et~al.]{Abbott2016}
Abbott, B.P.; Abbott, R.; Abbott, T.D.; Abernathy, M.R.; Acernese, F.; Ackley,
  K.; Adams, C.; Adams, T.; Addesso, P.; Adhikari, R.X.;  et~al.
\newblock Observation of Gravitational Waves from a Binary Black Hole Merger.
\newblock {\em Phys. Rev. Lett.} {\bf 2016}, {\em 116},~061102.
\newblock {\url{https://doi.org/10.1103/PhysRevLett.116.061102}}.

\bibitem[Abbott et~al.(2017)Abbott, Abbott, Abbott, Acernese, Ackley, Adams,
  Adams, Addesso, Adhikari, Adya, et~al.]{Abbott2017}
Abbott, B.P.; Abbott, R.; Abbott, T.D.; Acernese, F.; Ackley, K.; Adams, C.;
  Adams, T.; Addesso, P.; Adhikari, R.X.; Adya, V.B.;  et~al.
\newblock GW170817: Observation of Gravitational Waves from a Binary Neutron
  Star Inspiral.
\newblock {\em Phys. Rev. Lett.} {\bf 2017}, {\em 119},~161101.
\newblock {\url{https://doi.org/10.1103/PhysRevLett.119.161101}}.

\bibitem[Miller and Yunes(2019)]{Miller2019}
Miller, M.C.; Yunes, N.
\newblock The new frontier of gravitational waves.
\newblock {\em Nature} {\bf 2019}, {\em 568},~469--476.
\newblock {\url{https://doi.org/10.1038/s41586-019-1129-z}}.

\bibitem[Kornfeld et~al.(2019)Kornfeld, Arnold, Gross, Dahya, Klipstein, Gath,
  and Bettadpur]{Kornfeld2019}
Kornfeld, R.P.; Arnold, B.W.; Gross, M.A.; Dahya, N.T.; Klipstein, W.M.; Gath,
  P.F.; Bettadpur, S.
\newblock {GRACE}-{FO}: The Gravity Recovery and Climate Experiment Follow-On
  Mission.
\newblock {\em Journal of Spacecraft and Rockets} {\bf 2019}, {\em
  56},~931--951.
\newblock {\url{https://doi.org/10.2514/1.a34326}}.

\bibitem[Abich et~al.(2019)Abich, Abramovici, Amparan, Baatzsch, Okihiro, Barr,
  Bize, Bogan, Braxmaier, Burke, Clark, Dahl, Dahl, Danzmann, Davis, de~Vine,
  Dickson, Dubovitsky, Eckardt, Ester, Barranco, Flatscher, Flechtner, Folkner,
  Francis, Gilbert, Gilles, Gohlke, Grossard, Guenther, Hager, Hauden, Heine,
  Heinzel, Herding, Hinz, Howell, Katsumura, Kaufer, Klipstein, Koch, Kruger,
  Larsen, Lebeda, Lebeda, Leikert, Liebe, Liu, Lobmeyer, Mahrdt, Mangoldt,
  McKenzie, Misfeldt, Morton, M\"{u}ller, Murray, Nguyen, Nicklaus, Pierce,
  Ravich, Reavis, Reiche, Sanjuan, Sch\"{u}tze, Seiter, Shaddock, Sheard,
  Sileo, Spero, Spiers, Stede, Stephens, Sutton, Trinh, Voss, Wang, Wang, Ware,
  Wegener, Windisch, Woodruff, Zender, and Zimmermann]{Abich2019}
Abich, K.; Abramovici, A.; Amparan, B.; Baatzsch, A.; Okihiro, B.B.; Barr,
  D.C.; Bize, M.P.; Bogan, C.; Braxmaier, C.; Burke, M.J.;  et~al.
\newblock In-Orbit Performance of the {GRACE} Follow-On Laser Ranging
  Interferometer.
\newblock {\em Physical Review Letters} {\bf 2019}, {\em 123}.
\newblock {\url{https://doi.org/10.1103/physrevlett.123.031101}}.

\bibitem[Velicogna et~al.(2020)Velicogna, Mohajerani, A, Landerer, Mouginot,
  Noel, Rignot, Sutterley, Broeke, Wessem, and Wiese]{Velicogna2020}
Velicogna, I.; Mohajerani, Y.; A, G.; Landerer, F.; Mouginot, J.; Noel, B.;
  Rignot, E.; Sutterley, T.; Broeke, M.; Wessem, M.;  et~al.
\newblock Continuity of Ice Sheet Mass Loss in Greenland and Antarctica From
  the {GRACE} and {GRACE} Follow-On Missions.
\newblock {\em Geophysical Research Letters} {\bf 2020}, {\em 47}.
\newblock {\url{https://doi.org/10.1029/2020gl087291}}.

\bibitem[Landerer et~al.(2020)Landerer, Flechtner, Save, Webb, Bandikova,
  Bertiger, Bettadpur, Byun, Dahle, Dobslaw, Fahnestock, Harvey, Kang,
  Kruizinga, Loomis, McCullough, Murb\"{o}ck, Nagel, Paik, Pie, Poole,
  Strekalov, Tamisiea, Wang, Watkins, Wen, Wiese, and Yuan]{Landerer2020}
Landerer, F.W.; Flechtner, F.M.; Save, H.; Webb, F.H.; Bandikova, T.; Bertiger,
  W.I.; Bettadpur, S.V.; Byun, S.H.; Dahle, C.; Dobslaw, H.;  et~al.
\newblock Extending the Global Mass Change Data Record: {GRACE} Follow-On
  Instrument and Science Data Performance.
\newblock {\em Geophysical Research Letters} {\bf 2020}, {\em 47}.
\newblock {\url{https://doi.org/10.1029/2020gl088306}}.

\bibitem[Flechtner et~al.(2015)Flechtner, Neumayer, Dahle, Dobslaw, Fagiolini,
  Raimondo, and G\"{u}ntner]{Flechtner2015}
Flechtner, F.; Neumayer, K.H.; Dahle, C.; Dobslaw, H.; Fagiolini, E.; Raimondo,
  J.C.; G\"{u}ntner, A.
\newblock What Can be Expected from the {GRACE}-{FO} Laser Ranging
  Interferometer for Earth Science Applications?
\newblock {\em Surveys in Geophysics} {\bf 2015}, {\em 37},~453--470.
\newblock {\url{https://doi.org/10.1007/s10712-015-9338-y}}.

\bibitem[Wegener et~al.(2020)Wegener, M\"{u}ller, Heinzel, and
  Misfeldt]{Wegener2020}
Wegener, H.; M\"{u}ller, V.; Heinzel, G.; Misfeldt, M.
\newblock Tilt-to-Length Coupling in the {GRACE} Follow-On Laser Ranging
  Interferometer.
\newblock {\em Journal of Spacecraft and Rockets} {\bf 2020}, {\em
  57},~1362--1372.
\newblock {\url{https://doi.org/10.2514/1.a34790}}.

\bibitem[Armano et~al.(2016)Armano, Audley, Auger, Baird, Bassan, Binetruy,
  Born, Bortoluzzi, Brandt, Caleno, Carbone, Cavalleri, Cesarini, Ciani,
  Congedo, Cruise, Danzmann, de~Deus~Silva, De~Rosa, Diaz-Aguil\'o, Di~Fiore,
  Diepholz, Dixon, Dolesi, Dunbar, Ferraioli, Ferroni, Fichter, Fitzsimons,
  Flatscher, Freschi, Garc\'{\i}a~Mar\'{\i}n, Garc\'{\i}a~Marirrodriga, Gerndt,
  Gesa, Gibert, Giardini, Giusteri, Guzm\'an, Grado, Grimani, Grynagier,
  Grzymisch, Harrison, Heinzel, Hewitson, Hollington, Hoyland, Hueller,
  Inchausp\'e, Jennrich, Jetzer, Johann, Johlander, Karnesis, Kaune, Korsakova,
  Killow, Lobo, Lloro, Liu, L\'opez-Zaragoza, Maarschalkerweerd, Mance,
  Mart\'{\i}n, Martin-Polo, Martino, Martin-Porqueras, Madden, Mateos,
  McNamara, Mendes, Mendes, Monsky, Nicolodi, Nofrarias, Paczkowski,
  Perreur-Lloyd, Petiteau, Pivato, Plagnol, Prat, Ragnit, Ra\"{\i}s,
  Ramos-Castro, Reiche, Robertson, Rozemeijer, Rivas, Russano, Sanju\'an,
  Sarra, Schleicher, Shaul, Slutsky, Sopuerta, Stanga, Steier, Sumner, Texier,
  Thorpe, Trenkel, Tr\"obs, Tu, Vetrugno, Vitale, Wand, Wanner, Ward, Warren,
  Wass, Wealthy, Weber, Wissel, Wittchen, Zambotti, Zanoni, Ziegler, and
  Zweifel]{Armano2016}
Armano, M.; Audley, H.; Auger, G.; Baird, J.T.; Bassan, M.; Binetruy, P.; Born,
  M.; Bortoluzzi, D.; Brandt, N.; Caleno, M.;  et~al.
\newblock Sub-Femto-$g$ Free Fall for Space-Based Gravitational Wave
  Observatories: LISA Pathfinder Results.
\newblock {\em Phys. Rev. Lett.} {\bf 2016}, {\em 116},~231101.
\newblock {\url{https://doi.org/10.1103/PhysRevLett.116.231101}}.

\bibitem[Armano et~al.(2021)Armano, Audley, Baird, Binetruy, Born, Bortoluzzi,
  Brandt, Castelli, Cavalleri, Cesarini, et~al.]{Armano2021}
Armano, M.; Audley, H.; Baird, J.; Binetruy, P.; Born, M.; Bortoluzzi, D.;
  Brandt, N.; Castelli, E.; Cavalleri, A.; Cesarini, A.;  et~al.
\newblock Sensor Noise in {LISA} Pathfinder: In-Flight Performance of the
  Optical Test Mass Readout.
\newblock {\em Physical Review Letters} {\bf 2021}, {\em 126},~131103.
\newblock {\url{https://doi.org/10.1103/physrevlett.126.131103}}.

\bibitem[Armano et~al.(2022)Armano, Audley, Baird, Binetruy, Born, Bortoluzzi,
  Brandt, Castelli, Cavalleri, Cesarini, Cruise, Danzmann, de~Deus~Silva,
  Diepholz, Dixon, Dolesi, Ferraioli, Ferroni, Fitzsimons, Flatscher, Freschi,
  Garc\'{\i}a, Gerndt, Gesa, Giardini, Gibert, Giusteri, Grimani, Grzymisch,
  Guzman, Harrison, Hartig, Hechenblaikner, Heinzel, Hewitson, Hollington,
  Hoyland, Hueller, Inchausp\'e, Jennrich, Jetzer, Johann, Johlander, Karnesis,
  Kaune, Killow, Korsakova, Lobo, L\'opez-Zaragoza, Maarschalkerweerd, Mance,
  Mart\'{\i}n, Martin-Polo, Martin-Porqueras, Martino, McNamara, Mendes,
  Mendes, Meshksar, Monsky, Nofrarias, Paczkowski, Perreur-Lloyd, Petiteau,
  Plagnol, Ramos-Castro, Reiche, Rivas, Robertson, Russano, Sanjuan, Slutsky,
  Sopuerta, Steier, Sumner, Texier, Thorpe, Vetrugno, Vitale, Wand, Wanner,
  Ward, Wass, Weber, Wissel, Wittchen, and Zweifel]{Armano2022}
Armano, M.; Audley, H.; Baird, J.; Binetruy, P.; Born, M.; Bortoluzzi, D.;
  Brandt, N.; Castelli, E.; Cavalleri, A.; Cesarini, A.;  et~al.
\newblock Sensor noise in LISA Pathfinder: An extensive in-flight review of the
  angular and longitudinal interferometric measurement system.
\newblock {\em Phys. Rev. D} {\bf 2022}, {\em 106},~082001.
\newblock {\url{https://doi.org/10.1103/PhysRevD.106.082001}}.

\bibitem[Weber et~al.(2022)Weber, Bortoluzzi, Bosetti, Consolini, Dolesi, and
  Vitale]{Weber2022}
Weber, W.J.; Bortoluzzi, D.; Bosetti, P.; Consolini, G.; Dolesi, R.; Vitale, S.
\newblock Application of {LISA} Gravitational Reference Sensor Hardware to
  Future Intersatellite Geodesy Missions.
\newblock {\em Remote Sensing} {\bf 2022}, {\em 14},~3092.
\newblock {\url{https://doi.org/10.3390/rs14133092}}.

\bibitem[Drever et~al.(1983)Drever, Hall, Kowalski, Hough, Ford, Munley, and
  Ward]{Drever1983}
Drever, R.W.P.; Hall, J.L.; Kowalski, F.V.; Hough, J.; Ford, G.M.; Munley,
  A.J.; Ward, H.
\newblock Laser phase and frequency stabilization using an optical resonator.
\newblock {\em Applied Physics B Photophysics and Laser Chemistry} {\bf 1983},
  {\em 31},~97--105.
\newblock {\url{https://doi.org/10.1007/bf00702605}}.

\bibitem[Webster et~al.(2007)Webster, Oxborrow, and Gill]{Webster2007}
Webster, S.A.; Oxborrow, M.; Gill, P.
\newblock {Vibration insensitive optical cavity}.
\newblock {\em Phys. Rev. A - At. Mol. Opt. Phys.} {\bf 2007}, {\em 75},~1--4.
\newblock {\url{https://doi.org/10.1103/PhysRevA.75.011801}}.

\bibitem[Webster et~al.(2008)Webster, Oxborrow, Pugla, Millo, and
  Gill]{Webster2008}
Webster, S.A.; Oxborrow, M.; Pugla, S.; Millo, J.; Gill, P.
\newblock {Thermal-noise-limited optical cavity}.
\newblock {\em Physical Review A} {\bf 2008}, {\em 77},~033847--6.
\newblock {\url{https://doi.org/10.1103/PhysRevA.77.033847}}.

\bibitem[Webster and Gill(2011)]{Webster2011}
Webster, S.; Gill, P.
\newblock {Force-insensitive optical cavity}.
\newblock {\em Optics Letters} {\bf 2011}, {\em 36},~3572.
\newblock {\url{https://doi.org/10.1364/OL.36.003572}}.

\bibitem[Arie et~al.(1992)Arie, Schiller, Gustafson, and Byer]{Arie1992}
Arie, A.; Schiller, S.; Gustafson, E.K.; Byer, R.L.
\newblock Absolute frequency stabilization of diode-laser-pumped Nd:{YAG}
  lasers to hyperfine transitions in molecular iodine.
\newblock {\em Optics Letters} {\bf 1992}, {\em 17},~1204.
\newblock {\url{https://doi.org/10.1364/ol.17.001204}}.

\bibitem[Leonhardt and Camp(2006)]{Leonhardt2006}
Leonhardt, V.; Camp, J.B.
\newblock Space interferometry application of laser frequency stabilization
  with molecular iodine.
\newblock {\em Applied Optics} {\bf 2006}, {\em 45},~4142.
\newblock {\url{https://doi.org/10.1364/ao.45.004142}}.

\bibitem[Gerberding et~al.(2017)Gerberding, Isleif, Mehmet, Danzmann, and
  Heinzel]{Gerberding2017}
Gerberding, O.; Isleif, K.S.; Mehmet, M.; Danzmann, K.; Heinzel, G.
\newblock Laser-Frequency Stabilization via a Quasimonolithic Mach-Zehnder
  Interferometer with Arms of Unequal Length and Balanced dc Readout.
\newblock {\em Physical Review Applied} {\bf 2017}, {\em 7},~024027.
\newblock {\url{https://doi.org/10.1103/physrevapplied.7.024027}}.

\bibitem[Huarcaya et~al.(2023)Huarcaya, \'Alvarez, Penkert, Gozzo, Cano,
  Yamamoto, Delgado, Mehmet, Danzmann, and Heinzel]{bib:VH}
Huarcaya, V.; \'Alvarez, M.D.; Penkert, D.; Gozzo, S.; Cano, P.M.; Yamamoto,
  K.; Delgado, J.J.E.; Mehmet, M.; Danzmann, K.; Heinzel, G.
\newblock $2\ifmmode\times\else\texttimes\fi{}{10}^{\ensuremath{-}13}$
  Fractional Laser-Frequency Stability with a 7-cm Unequal-Arm Mach-Zehnder
  Interferometer.
\newblock {\em Phys. Rev. Appl.} {\bf 2023}, {\em 20},~024078.
\newblock {\url{https://doi.org/10.1103/PhysRevApplied.20.024078}}.

\bibitem[Isleif(2018)]{bib:katha}
Isleif, K.S.
\newblock {\em Laser interferometry for LISA and satellite geodesy missions};
  Hannover : Institutional Repository of Leibniz Universität Hannover,  2018.
\newblock {\url{https://doi.org/10.15488/3526}}.

\bibitem[Yang et~al.(2020)Yang, Yamamoto, Huarcaya, Vorndamme, Penkert,
  Fernández~Barranco, Schwarze, Mehmet, Esteban~Delgado, Jia, Heinzel, and
  Dovale~\'Alvarez]{bib:SEDI}
Yang, Y.; Yamamoto, K.; Huarcaya, V.; Vorndamme, C.; Penkert, D.;
  Fernández~Barranco, G.; Schwarze, T.S.; Mehmet, M.; Esteban~Delgado, J.J.;
  Jia, J.;  et~al.
\newblock Single-Element Dual-Interferometer for Precision Inertial Sensing.
\newblock {\em Sensors} {\bf 2020}, {\em 20}.
\newblock {\url{https://doi.org/10.3390/s20174986}}.

\bibitem[Gerberding(2015)]{bib:DFM}
Gerberding, O.
\newblock Deep frequency modulation interferometry.
\newblock {\em Opt. Express} {\bf 2015}, {\em 23},~14753--14762.
\newblock {\url{https://doi.org/10.1364/OE.23.014753}}.

\bibitem[Heinzel et~al.(2010)Heinzel, Cervantes, Mar\'{i}n, Kullmann, Feng, and
  Danzmann]{bib:DPM}
Heinzel, G.; Cervantes, F.G.; Mar\'{i}n, A.F.G.; Kullmann, J.; Feng, W.;
  Danzmann, K.
\newblock Deep phase modulation interferometry.
\newblock {\em Opt. Express} {\bf 2010}, {\em 18},~19076--19086.
\newblock {\url{https://doi.org/10.1364/OE.18.019076}}.

\bibitem[Dovale-{\'{A}}lvarez(2019)]{Dovale2019PhD}
Dovale-{\'{A}}lvarez, M.
\newblock {\em Optical Cavities for Optical Atomic Clocks, Atom Interferometry
  and Gravitational-Wave Detection}; Springer International Publishing,  2019.
\newblock {\url{https://doi.org/10.1007/978-3-030-20863-9}}.

\bibitem[Instruments()]{bib:moku}
Instruments, L.
\newblock Liquid Instruments.
\newblock \url{https://www.liquidinstruments.com}.

\bibitem[Greenhall and Riley(2003)]{Greenhall2003}
Greenhall, C.A.; Riley, W.J.
\newblock {Uncertainty of stability variances based on finite differences} {\bf
  2003}.
\newblock {\url{https://doi.org/2014/38105}}.

\bibitem[Tr\"{o}bs and Heinzel(2006)]{Troebs2006}
Tr\"{o}bs, M.; Heinzel, G.
\newblock Improved spectrum estimation from digitized time series on a
  logarithmic frequency axis.
\newblock {\em Measurement} {\bf 2006}, {\em 39},~120--129.
\newblock {\url{https://doi.org/10.1016/j.measurement.2005.10.010}}.

\bibitem[Riley and Howe(2008)]{NIST1065}
Riley, W.; Howe, D.
\newblock {Handbook of Frequency Stability Analysis}.
\newblock {\em NIST Special Publication - 1065} {\bf 2008}.

\bibitem[Watchi et~al.(2018)Watchi, Cooper, Ding, Mow-Lowry, and
  Collette]{Watchi2018}
Watchi, J.; Cooper, S.; Ding, B.; Mow-Lowry, C.M.; Collette, C.
\newblock {Contributed Review: A review of compact interferometers}.
\newblock {\em Review of Scientific Instruments} {\bf 2018}, {\em 89},~121501,
  \href{http://xxx.lanl.gov/abs/https://pubs.aip.org/aip/rsi/article-pdf/doi/10.1063/1.5052042/15603194/121501\_1\_online.pdf}{{\normalfont
  [https://pubs.aip.org/aip/rsi/article-pdf/doi/10.1063/1.5052042/15603194/121501\_1\_online.pdf]}}.
\newblock {\url{https://doi.org/10.1063/1.5052042}}.

\bibitem[Pisani and Zucco(2018)]{Pisani2018}
Pisani, M.; Zucco, M.
\newblock An accelerometer for spaceborne application with interferometric
  readout.
\newblock {\em Measurement} {\bf 2018}, {\em 122},~507--512.
\newblock
  {\url{https://doi.org/https://doi.org/10.1016/j.measurement.2018.03.014}}.

\bibitem[Isleif et~al.(2019)Isleif, Heinzel, Mehmet, and
  Gerberding]{Isleif2019}
Isleif, K.S.; Heinzel, G.; Mehmet, M.; Gerberding, O.
\newblock Compact Multifringe Interferometry with Subpicometer Precision.
\newblock {\em Phys. Rev. Appl.} {\bf 2019}, {\em 12},~034025.
\newblock {\url{https://doi.org/10.1103/PhysRevApplied.12.034025}}.

\bibitem[Yan et~al.(2022)Yan, Yeh, and Mao]{Yan2022}
Yan, H.; Yeh, H.C.; Mao, Q.
\newblock High precision six-degree-of-freedom interferometer for test mass
  readout.
\newblock {\em Classical and Quantum Gravity} {\bf 2022}, {\em 39},~075024.
\newblock {\url{https://doi.org/10.1088/1361-6382/ac5923}}.

\bibitem[Zhang and Guzman(2022)]{Zhang2022}
Zhang, Y.; Guzman, F.
\newblock Quasi-monolithic heterodyne laser interferometer for inertial
  sensing.
\newblock {\em Opt. Lett.} {\bf 2022}, {\em 47},~5120--5123.
\newblock {\url{https://doi.org/10.1364/OL.473476}}.

\bibitem[Smetana et~al.(2022)Smetana, Walters, Bauchinger, Ubhi, Cooper,
  Hoyland, Abbott, Baune, Fritchel, Gerberding, K\"ohnke, Miao, Rode, and
  Martynov]{Smetana2022}
Smetana, J.; Walters, R.; Bauchinger, S.; Ubhi, A.S.; Cooper, S.; Hoyland, D.;
  Abbott, R.; Baune, C.; Fritchel, P.; Gerberding, O.;  et~al.
\newblock Compact Michelson Interferometers with Subpicometer Sensitivity.
\newblock {\em Phys. Rev. Appl.} {\bf 2022}, {\em 18},~034040.
\newblock {\url{https://doi.org/10.1103/PhysRevApplied.18.034040}}.

\bibitem[Kranzhoff et~al.(2022)Kranzhoff, Lehmann, Kirchhoff, Carlassara,
  Cooper, Koch, Leavey, Lück, Mow-Lowry, Wöhler, von Wrangel, and
  Wu]{Kranzhoff2022}
Kranzhoff, S.L.; Lehmann, J.; Kirchhoff, R.; Carlassara, M.; Cooper, S.J.;
  Koch, P.; Leavey, S.; Lück, H.; Mow-Lowry, C.M.; Wöhler, J.;  et~al.
\newblock A vertical inertial sensor with interferometric readout.
\newblock {\em Classical and Quantum Gravity} {\bf 2022}, {\em 40},~015007.
\newblock {\url{https://doi.org/10.1088/1361-6382/aca45b}}.

\end{thebibliography}

\PublishersNote{}
\end{document}